\documentclass[twocolumn,pra,10pt,aps,nofootinbib]{revtex4-1}
\usepackage{relsize}
\usepackage{upgreek}

\usepackage{amsmath,amssymb,graphicx}
\usepackage{mathtools}
\usepackage{array,booktabs}
\usepackage{matlab-prettifier}
\usepackage{listings}
\usepackage{bm}
\newcommand{\pd}{\partial}
\def\d{\mathrm{d}}
\renewcommand{\b}{\boldsymbol}
\def\b#1{\mathbf{#1}}
\newcommand{\cO}{\mathcal{O}}
\begin{document}
\title{Quasistatic limit of the electric-magnetic coupling blocks of the $T$-matrix for spheroids}

\date{\today}
\author{Matt R. A. Maji\'c}
\email{mattmajic@gmail.com}
\author{Eric C. Le Ru}
\email{eric.leru@vuw.ac.nz}
\affiliation{The MacDiarmid Institute for Advanced Materials and Nanotechnology,
School of Chemical and Physical Sciences, Victoria University of Wellington,
PO Box 600, Wellington 6140, New Zealand}

\begin{abstract}
The $T$-matrix formally describes the solution of any electromagnetic scattering problem by a given particle in a given medium at a given wavelength.
As such it is commonly used in a number of contexts, for example to predict the orientation-averaged optical properties of non-spherical particles.
The $T$-matrix for electromagnetic scattering can be divided into four blocks corresponding physically to coupling
between either magnetic or electric multipolar fields. Analytic expressions were recently derived for the electrostatic limit of the electric-electric $T$-matrix block $\b T^{22}$, of prolate spheroids. In such an electrostatic approximation, all the other blocks were zero. We here analyse the long-wavelength limit for the other blocks ($\b T^{21}$, $\b T^{12}$, $\b T^{11}$) corresponding to electric-magnetic, magnetic-electric, and magnetic-magnetic coupling respectively.
Analytic expressions (finite sums) are obtained in the case of spheroidal particles by expressing the fields with solutions to Laplace's equation, expanding the fields in terms of spheroidal harmonics and applying the boundary conditions. Similar expressions are also presented for the auxiliary matrices in the extended boundary condition method, often used in conjunction with the $T$-matrix formalism.
\end{abstract}

\maketitle

\section{Introduction}

The $T$-matrix is a widely used semi-analytic technique for the study of electromagnetic scattering by particles \cite{waterman1965matrix,waterman1971symmetry,1975BarberAO,1990Barber,1999DoicuJOSAA,Mishchenko2002,2004Database,2017Database}. In this approach, the electric and magnetic fields are expanded as series of vector spherical wavefunctions, and the $T$-matrix defines the linear relationship between the expansion coefficients of the incident and scattered fields. The $T$-matrix can be computed in many ways but a common approach is the extended boundary condition method (EBCM) \cite{2002Mishchenko}, which involves the division of two matrices whose matrix elements are given by integrals on the particle surface. It may also be possible to obtain the $T$-matrix directly from solving the problem from the boundary problem, which can be used to deduce analytic results for particles of simple shapes, for example spheroidal vector wave functions have been used to calculate the $T$-matrix for a spheroid \cite{schulz1998scattering}. This is the approach we apply here, but in the long-wavelength limit.

This manuscript follows from Ref.~\cite{TmatESA2017}, where analytic expressions were obtained for the long-wavelength limit of the $T$-matrix block for electric-electric multipole coupling, $\b T^{22}$, of a prolate spheroid. This limit is then equivalent to the solution of an electrostatic (or quasistatic) problem. The matrix elements were determined by solving the corresponding boundary problem (involving Laplace's equation) using spheroidal harmonics, and applying the expansions relating spherical/spheroidal harmonics \cite{Jansen2000,Antonov2002} to express the scattered field in terms of spherical harmonics and from there extract the entries of the $T$-matrix. Here we extend the approach to the other blocks governing interactions between electric and magnetic multipoles. The problem is more complicated as it can no longer be reduced to simply solving Laplace's equation, but similar analytic results can still be found. 
Note that unlike for $\b T^{22}$, where the quasistatic limit coincides with an exact electrostatics problem, the quasistatic limits of the other blocks are only physically meaningful as approximations of the corresponding time harmonic problem. This analytic limit may nevertheless be useful for fundamental studies of the $T$-matrix
method, for example in investigations of its convergence \cite{JQSRT2015} or related problems associated with the Rayleigh hypothesis \cite{2016AuguieJO}. It may also be used as a substitute for high order elements in cases when they are well approximated by their lowest order approximation \cite{allardice2014convergence}.

The manuscript is organized as follows. Section \ref{secT} is a brief recap of the $T$-matrix formalism and of the main notations. Section \ref{secT22} summarizes the results obtained in Ref.~\cite{TmatESA2017} for $\b T^{22}$. Section \ref{secT21} modifies the approach of Ref.~\cite{TmatESA2017} to obtain the quasistatic limit of $\b T^{21}, \b T^{12}$ for general axisymmetric particles and derive analytic expressions for the matrix elements. Section \ref{secT11} further modifies the approach to obtain $\b T^{11}$. Section \ref{disc} discusses these results, in particular proposes a definition for the generalized depolarization factors for spheroids, and exploits these to discuss the multipolar plasmon resonance conditions for metallic spheroids.

\section{General approach/notations}
\label{secT}
\subsection{T-matrix formalism}
We first summarize the $T$-matrix formalism for electromagnetic scattering \cite{2002Mishchenko}. We consider some known time-harmonic external electromagnetic field $\b E_e, \b H_e$ incident on a non-magnetic particle in a non-absorbing medium. A time dependence $e^{-i\omega t}$ is implied. The permittivity inside and outside the particle are denoted $\epsilon_i$ , $\epsilon_o$, with their ratio $\epsilon=\epsilon_i/\epsilon_o$ and relative refractive index $s=\sqrt{\epsilon}$ (possibly complex and wavelength dependent).
The wavenumber inside and outside the particle are denoted $k_i$ and $k_o$ with $k_i=sk_o$.
The external field creates an internal field $\b E_i, \b H_i$ inside the scatterer and a scattered field $\b E_s, \b H_s$, so that the field outside the particle is $\b E_o=\b E_e+\b E_s,~~ \b H_o=\b H_e +\b H_s$. The fields are expanded as series of vector spherical wave functions:
\begin{align}
\b E_e &= E_0 \sum_{n,m} a_n^m \b {RgM}_n^m(k_o\b r) + b_n^m \b {RgN}_n^m(k_o\b r), \\
\b H_e &= H_0 \sum_{n,m} a_n^m \b {RgN}_n^m(k_o\b r) + b_n^m \b {RgM}_n^m(k_o\b r), \\
\b E_i &= E_0 \sum_{n,m} c_n^m \b {RgM}_n^m(k_i\b r) + d_n^m \b {RgN}_n^m(k_i\b r), \\
\b H_i &= H_0s \sum_{n,m} c_n^m \b {RgN}_n^m(k_i\b r) + d_n^m \b {RgM}_n^m(k_i\b r), \\
\b E_s &= E_0 \sum_{n,m} p_n^m \b {M}_n^m(k_o\b r) + q_n^m \b {N}_n^m(k_o\b r), \\
\b H_s &= H_0 \sum_{n,m} p_n^m \b {N}_n^m(k_o\b r) + q_n^m \b {M}_n^m(k_o\b r), 
\end{align} 
where $E_0$ is the incident electric field strength and
$H_0= E_0 k_o/(i\omega\mu_0)$.  $\b {RgM}$, $\b {RgN}$ denote the regular wavefunctions while $\b M$, $\b N$ denote the singular wavefunctions, corresponding to magnetic and electric multipolar fields, respectively. Our definitions differ by $(-)^m$ to those in appendix C of \cite{Mishchenko2002}.

The problem is to determine the coefficients $c_n^m, d_n^m, p_n^m, q_n^m$ that satisfy the boundary conditions at the surface of the scatterer
(with $\hat{\b n}$ the unit normal vector):
\begin{align}
\epsilon \hat{\b n} \cdot \b E_i &= \hat{\b n} \cdot \b E_o, \qquad \hat{\b n} \times \b E_i = \hat{\b n}\times\b E_o, \nonumber\\
\hat{\b n}\cdot \b H_i &= \hat{\b n}\cdot \b H_o, \qquad \hat{\b n}\times\b H_i = \hat{\b n}\times\b H_o,
\end{align}
and the Sommerfeld radiation condition for the scattered field at infinity.

By linearity of Maxwell's equations, the coefficients are related by linear expressions commonly expressed in matrix form as:
\begin{align}
\begin{bmatrix}
{\bf p} \\ {\bf q}
\end{bmatrix}
&=\begin{bmatrix}
\b T^{11} & \b T^{12} \\ \b T^{21} & \b T^{22}
\end{bmatrix}
\begin{bmatrix}
{\bf a} \\ {\bf b}
\end{bmatrix}
= \b T \begin{bmatrix}
{\bf a} \\ {\bf b}
\end{bmatrix},
\end{align}
which defines the $T$-matrix.
The column vectors ${\bf a}$, ${\bf b}$, ${\bf c}$, ${\bf d}$, ${\bf p}$ and ${\bf q}$ contain $a_n^m$, $b_n^m$ $c_n^m$, $d_n^m$, $p_n^m$ and $q_n^m$ as components, for all $n$ and $m$. 
Within the EBCM, one typically also defines the $P$- and $Q$-matrices as:
\begin{align}
\begin{bmatrix}
{\bf a} \\ {\bf b}
\end{bmatrix}
&=\begin{bmatrix}
\b Q^{11} & \b Q^{12} \\ \b Q^{21} & \b Q^{22}
\end{bmatrix}
\begin{bmatrix}
{\bf c} \\ {\bf d}
\end{bmatrix}, \\[0.2cm]
\begin{bmatrix}
{\bf p} \\ {\bf q}
\end{bmatrix}
&= -\begin{bmatrix}
\b P^{11} & \b P^{12} \\ \b P^{21} & \b P^{22}
\end{bmatrix}
\begin{bmatrix}
{\bf c} \\ {\bf d}
\end{bmatrix}. \label{P}
\end{align}
and we here also introduce the matrix $\b R=\b Q^{-1}$.
Note that Ref.~\cite{2002Mishchenko} uses ${\b{RgQ}}$ instead of $\b P$.

For axisymmetric particles (such as spheroids), a major simplification is that all matrices are decoupled for each $m$, and we may therefore treat each $m$ separately.
The matrix elements for a given $m$ will then be denoted $T^{ij}_{nk|m}$.

Moreover, for particles with reflection symmetry with respect to the $z=0$ plane (like spheroids), half of the matrix elements are zero, namely:
\begin{align}
A^{11}_{nk}&=A^{22}_{nk}=0\quad n+k \text{ odd}, \\
A^{21}_{nk}&=A^{12}_{nk}=0\quad  n+k \text{ even},
\end{align}
for $A=P,Q,R,T$.

\subsection{Spheroidal coordinates and harmonics}
\begin{figure}[b]
\includegraphics[scale=.26]{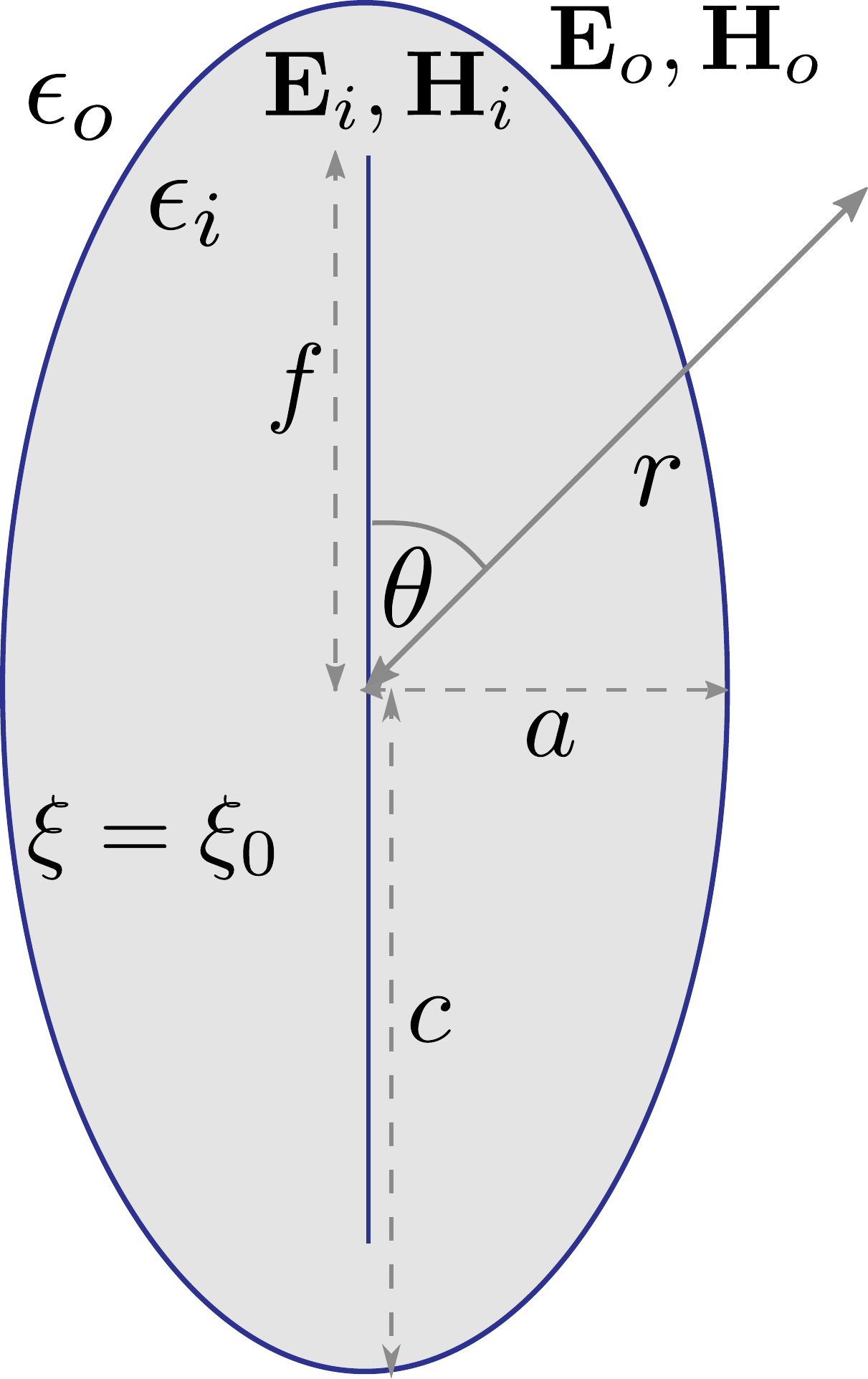}
\includegraphics[scale=.29]{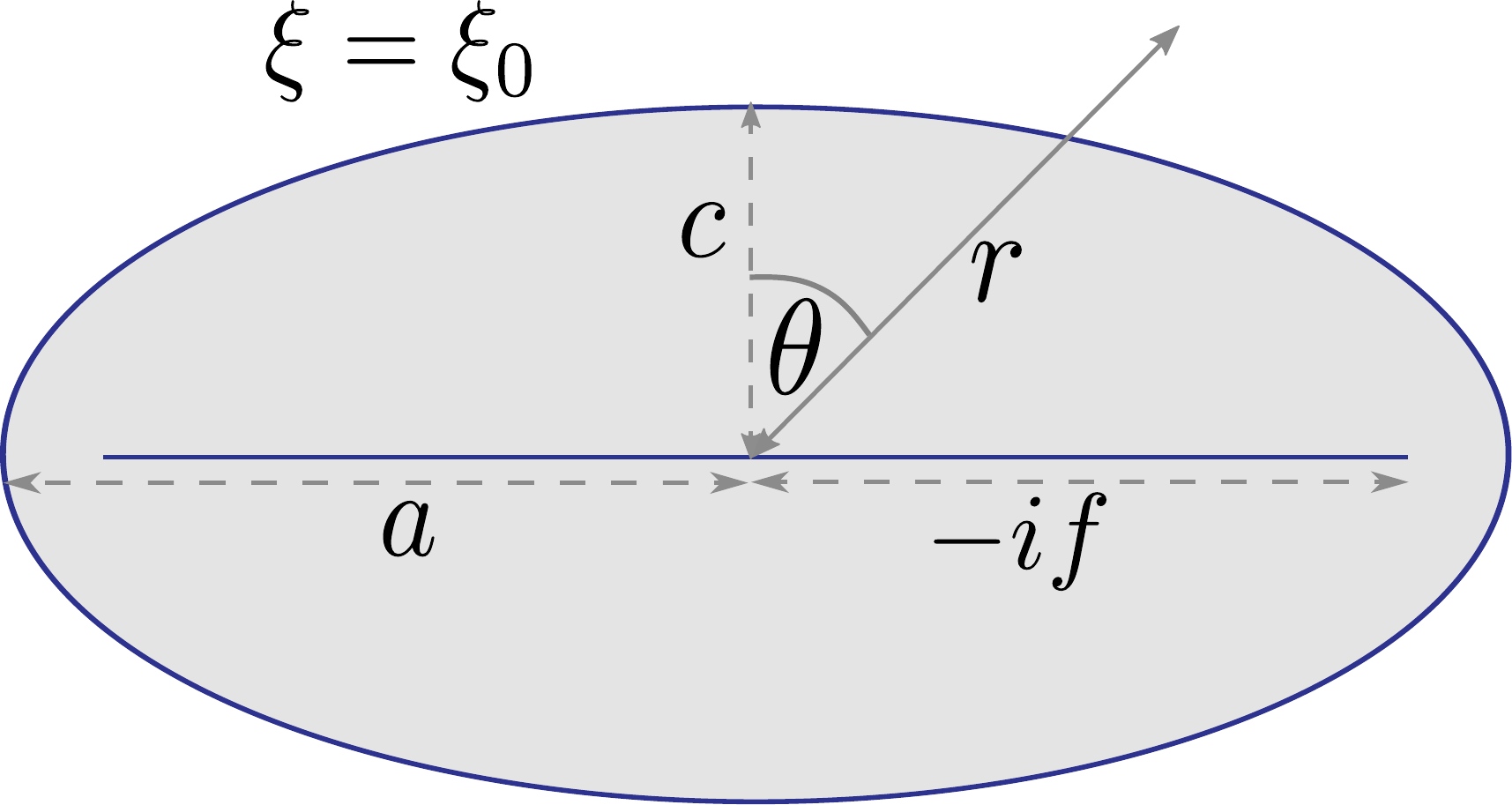}
\caption{Schematics of the scattering problem for prolate and oblate  spheroids.}
\end{figure}
We consider a dielectric spheroid (prolate or oblate) of semi-height $c$ along the $z$-axis and semi-width $a$ along $x,y$. It will be convenient to define oblate spheroidal coordinates and parameters by exactly the same formulae, since this choice means the $T$-matrix expressions for prolate and oblate spheroids will also have exactly the same expressions. We define the focal parameter $f=\sqrt{c^2-a^2}$. Then prolate spheroids have $c>a$, half focal-length $f$, and oblate spheroids have $a>c$, focal disk radius $-if = \sqrt{a^2-c^2}$.
The spheroidal coordinates $\xi,\eta$ are defined in terms of $r^+,r^-$, the distance from the top and bottom focal points:
\begin{align}
&\xi=\frac{r^++r^-}{2f}, \qquad \eta=\frac{r^+-r^-}{2f} \nonumber\\
&r^\pm=\sqrt{r^2\pm2fr\cos\theta+f^2}.\label{xieta} 
\end{align}
for both prolate and oblate. $\xi=\xi_0=c/\sqrt{c^2-a^2}$ defines the surface of our scatterer. 
For prolate spheroidal coordinates, $\xi_0$ ranges from 1 (needle) to $\infty$ (large sphere), while for oblate coordinates, $\xi_0$ ranges from 0 (disk) to $-i\infty$ (large sphere). 

The solution of Laplace's equation in spheroidal harmonics involves the product of Legendre functions of the first or second kind $P_n^m(\xi)$, or $Q_n^m(\xi)$, with  $P_n^m(\eta)$. Their derivatives are denoted with a prime $'$.
Several similar definitions exist for these, and we will here use the ones specified in the Appendix, which ensure that all derived formulae are also correct for oblate spheroids.


\subsection{Quasistatic/long-wavelength approximation}

The term quasistatic here means that the wavelength of the light both inside and outside the particle is long in comparison to the particle size. 
As per Ref.~\cite{SpheroidApproximation2018} we shall define a size parameter $X=k_oc$, where for small particles relative to the wavelength, $X\ll 1$ and the fields may be considered as asymptotic expansions in powers of $X$, and here we consider just the lowest two orders. This approximation also requires $k_i c=sX\ll 1$, so the relative refractive index $s$ must not be too large. This definition for $X$ is convenient here, but as shown in Ref.~\cite{SpheroidApproximation2018}, it is usually more relevant to define the size parameter in terms of the radius of the sphere of equivalent volume.
One should also note that spherical Bessel functions of higher order $n$ can be well approximated by their dominant term (small $X$ approximation) for up to $X\lesssim n$ \cite{allardice2014convergence}, so one can expect that our final expressions for matrix elements for large $n,k$ will be valid for relatively large $X$.

In this analysis it is crucial to be aware of how the basis functions and matrix elements depend on $X$. In the near field, $X\approx k_or$: 
\begin{align}
\b {RgM}_{nm}^{(0)}&=\cO(X^n),     \qquad &\b M_{nm}^{(0)}=\cO(X^{-n-1}), \label{EqnOrderM}\\
\b {Rg N}_{nm}^{(0)}&=\cO(X^{n-1}), &~\b N_{nm}^{(0)}=\cO( X^{-n-2}), \label{EqnOrderN}
\end{align}
and $E_0=\cO(X^0)$ while $H_0=\cO(X^1)$.
The quasistatic limit of the $Q$-matrix and $R$-matrix elements take a special form for a spheroidal particle \cite{JQSRT2012,JQSRT2013}
and the order of the dominant terms for all matrices are summarized below \cite{JQSRT2012,SpheroidApproximation2018}:
\begin{align}
Q^{11}_{nk} &= \cO ( X^{[n<k](k-n+2)}), \quad Q^{21}_{nk} = \cO ( X^{[n<k](k-n+1)}), \nonumber\\
Q^{12}_{nk} &= \cO ( X^{[n<k](k-n+1)}), \quad Q^{22}_{nk} = \cO ( X^{[n<k](k-n)}),
\label{EqnOrderQ}
\end{align}
where $[n<k] = 1$ if $n<k$ and 0 otherwise. For general axisymmetric particles the $[n<k]$ is not present. For $\b P$:
\begin{align}
P^{11}_{nk} &= \cO ( X^{k+n+3}),  \quad
&&P^{12}_{nk} = \cO ( X^{k+n+2}), \nonumber\\
P^{21}_{nk} &= \cO ( X^{k+n+2}),   \quad
&&P^{22}_{nk} = \cO ( X^{k+n+1}).
\label{EqnOrderP}
\end{align}
$\b T$ has identical behaviour as $\b P$, and $\b R$ as $\b Q$, so:
\begin{align}
R^{ij}_{nk} \propto Q^{ij}_{nk} 
~~\mathrm{and}~~ T^{ij}_{nk} \propto P^{ij}_{nk} 
\quad\mathrm{For~} i,j =1,2.\label{EqnOrderT}
\end{align}

The aim of this work is to find analytic expressions for these dominant terms for all matrix elements
for all matrix elements of $\b T$ for all $n,k$. In Ref.~\cite{TmatESA2017}, expressions were derived for the $\b T^{22}$ block, while
in Ref.~\cite{SpheroidApproximation2018}, expressions were obtained for all blocks but only for $n,k \leq 3$.

In the process, we will also derive matrix elements for $\b P$, $\b Q$, and $\b R$, except for the lower
triangular parts of $\b Q$ and $\b R$, because they reduce to zero in the limit of this problem.

\section{Summary for 22 blocks}
\label{secT22}

For completeness we summarize the quasistatic limit for the elements of $\b P^{22}$, $\b Q^{22}$, $\b R^{22}$, $\b T^{22}$.
These results were derived in Ref.~\cite{TmatESA2017} for $\b P$, $\b Q$, and $\b T$. They are here slightly rearranged and
the corresponding expression for $\b R$ is also given.
\begin{widetext}
\begin{align}	
P^{22(0)}_{nk|m} &= -i s^{k-1} (k_of)^{n+k+1}~B_n^m B_k^m (s^2-1)(\xi_0^2-1)(-)^m   \nonumber\\
	&\times\sum_{p=|m|}^{\min(n,k)} \frac{e_{nk}e_{np}(2p+1)~P_p^{-m}(\xi_0) P_p^{m\prime}(\xi_0)}{(n-p)!!(n+p+1)!!(k-p)!!(k+p+1)!!},  
\label{esP} \\[5pt]\nonumber\\
Q^{22(0)}_{nk|m} &=  s^{k-1}  \delta_{nk} 
+ s^{k-1}(k_of)^{k-n}~\frac{B_k^m}{B_n^m}(s^2-1)(\xi_0^2-1) \nonumber\\
 &\times\sum_{p=n}^k e_{nk}e_{np} \frac{(-)^{(p-n)/2}(2p+1)(n+p-1)!!}{(p-n)!!(k-p)!!(k+p+1)!!}  Q_p^{-m}(\xi_0) P_p^{m\prime}(\xi_0) ,
\label{esQ} 
\end{align}
\begin{align}
R^{22(0)}_{nk|m} &= s^{1-n}(k_of)^{k-n}~\frac{B_k^m}{B_n^m} \nonumber\\
 &\times\sum_{p=n}^k e_{nk}e_{np} \frac{(-)^{(p-n)/2}(2p+1)(n+p-1)!!}{(p-n)!!(k-p)!!(k+p+1)!!}\frac{1}{1+(s^2-1)L_p^m(\xi_0)} ,
\label{esR} \\[5pt]\nonumber\\
T^{22(0)}_{nk|m} &= i (k_of)^{n+k+1}~B_n^m B_k^m (s^2-1)(\xi_0^2-1)(-)^m \nonumber \\
	&\times\sum_{p=|m|}^{\min(n,k)} 	\frac{e_{nk}e_{np} (2p+1)}{(n-p)!!(n+p+1)!!(k-p)!!(k+p+1)!!}\frac{P_p^{-m}(\xi_0) P_p^{m\prime}(\xi_0)}{1+(s^2-1)L^m_p(\xi_0)} , 
\label{esT}
\end{align}
\end{widetext}

where
\begin{align}
B_n^m&=\frac{1}{(2n-1)!!}\sqrt{\frac{(n+1)(n+m)!(n-m)!}{n(2n+1)}}, \\
e_{nk}&=
\begin{cases} 
1 & n+k \text{ even} \\
0 & n+k \text{ odd}
\end{cases}.
\end{align}

We have also introduced the generalized depolarization factors:
\begin{align}
L^m_n&=(\xi_0^2-1)P_n^{m\prime}(\xi_0)Q_n^{-m}(\xi_0) 
\end{align}
which will be further discussed in Sec.~\ref{secLnm}.

While in Ref.~\cite{TmatESA2017} the spheroid was assumed to be prolate, these formulae also apply to oblate spheroids 
using the definitions of the Legendre functions given in the Appendix.

\section{Quasistatic limit of $\b T^{21}$, $\b T^{12}$}
\label{secT21}

\subsection{General approach}

In Ref.~\cite{TmatESA2017}, the quasistatic limit of $\b T^{22}$ was found by solving an equivalent electrostatics problem, where the Helmholtz equation reduces to Laplace's equation. Here $\b T^{21}$ (and $\b T^{12}$) are zero to lowest order in $X$, so the next order in $X$ must be considered. 

We focus on obtaining $\b T^{21}$, as $\b T^{12}$ can then be obtained through
\begin{align}
T^{12}_{nk|m}=-T^{21}_{kn|m}.
\end{align}

To extract the quasistatic limit of $\b T^{21}$, we consider a particular quasistatic excitation, consisting of only magnetic multipoles - that is
\begin{align}
\b E_e = E_0 \sum_{nm} a_n^m \b{RgM}^{(0)}_{nm}(k_o\b r) , \\
\b H_e = H_0 \sum_{nm} a_n^m \b{RgN}^{(0)}_{nm}(k_o\b r).
\end{align} 
The superscript $(0)$ denotes the lowest non-zero order in $X$.
Considering the orders of the spherical vector wavefunctions given in \eqref{EqnOrderM}-\eqref{EqnOrderN}, we will impose that the coefficients $a_n^m$ depend on $X$ as $a_n^m\propto X^{1-n}$, so that all multipole terms in the expansions are of the same order in $X$, and as a result we have $\b E_i,\b H_i=\cO(X^1)$, with every term in the sum being $\cO(X^1)$. It can then be shown that all elements of $\b T^{21}$ will be obtainable by reduction to the lowest non-zero order of $X$. A possible physical example where $a_n^m\propto X^{1-n}$, is a low frequency radiating magnetic dipole located outside the scatterer.  If we instead considered a plane wave excitation, only the lowest order multipoles would be non-negligible, and we would only obtain information about the low order entries of the scattering matrices. 

We now analyse which terms in the series of the incident and scattered fields can be neglected. 
 We also need to take into account the dependence of lowest order of $\b T^{21}$, $\b R^{21}$, which were given in Eqs~\ref{EqnOrderQ}--\ref{EqnOrderT}. 
The significant parts of the internal and scattered fields to $\cO(X^1)$ will be 
\begin{align}
\b E_i &= E_0\sum_{n,m} c_n^m \b {RgM}_{nm}^{(0)}(k_i\b r) + d_n^m \b {RgN}_{nm}^{(0)}(k_i\b r), \nonumber\\
\b H_i &= sH_0\sum_{n,m} c_n^m \b {RgN}_{nm}^{(0)}(k_i\b r) \nonumber\\
\b E_s &= E_0\sum_{n,m} q_n^m \b N_{nm}^{(0)}(k_o\b r), \nonumber\\
\b H_s &= \cO(X^2).
\end{align}
In the long-wavelength limit the magnetic field does not interact with the particle. The magnetic boundary conditions are therefore solved simply by setting the internal magnetic field identical to the external field, that is
\begin{align}
c_n^m=&s^{-n}a_n^m \nonumber\\
~\Leftrightarrow~ Q_{nk}^{11(0)}=\delta_{nk}s^n, \qquad &
 R_{nk}^{11(0)}=\delta_{nk}s^{-n}.\label{R11}
\end{align}
This also means that the magnetic-multipolar part of the electric field is equal to the magnetic-multipolar part of the incident electric field. However, this alone does not satisfy the electric boundary conditions, so the problem now is to solve for the coefficients $d_n^m$ and $q_n^m$, knowing both $a_n^m$ and $c_n^m$. For this problem the matrix relations between the known and unknown coefficients are
\begin{align}
\b q= \b T^{21} \b a, \qquad \b d= \b R^{21} \b a, \label{TR}
\end{align}

In the long-wavelength limit the vector spherical wave functions are
\begin{align}
\mathbf{RgM}_{nm}^{(0)} &= \gamma_n^m \frac{k^n}{(2n+1)!!}~\b r\times \nabla[r^nP_n^m(\cos\theta)e^{im\phi}], \label{limM1}\\ 
\vspace{.2cm}
\mathbf{RgN}_{nm}^{(0)} &=\gamma_n^m \frac{(n+1)k^{n-1}}{(2n+1)!!}~{\b \nabla}[r^nP_n^m(\cos\theta)e^{im\phi}], \label{limN1}\\ 
\vspace{.2cm}
\mathbf{M}_{nm}^{(0)} &= \gamma_n^m \frac{(2n-1)!!}{ik^{n+1}}~\b r\times \nabla[r^{-n-1}P_n^m(\cos\theta)e^{im\phi}], \label{limM3}\\ 
\vspace{.2cm}
\mathbf{N}_{nm}^{(0)} &= \gamma_n^m \frac{in(2n-1)!!}{k^{n+2}}{\b \nabla} [r^{-n-1}P_n^m(\cos\theta)e^{im\phi}], \label{limN3}\\
\mathrm{where~} & \gamma_n^m=\sqrt{\frac{2n+1}{4\pi n(n+1)}\frac{(n-m)!}{(n+m)!}}.
\end{align}
This means we can express the electric field as:
\begin{align}
\b E_e&=\b r\times \nabla U, \label{E pot1} \\
\b E_i&=\b r\times \nabla U - \nabla V_i,  \label{E pot2} \\
\b E_s&= -\nabla V_s,  \label{E pot3} \\
\text{with }& \nabla^2 U=\nabla^2 V_i=\nabla^2 V_s=0. \label{E pot4} 
\end{align}
Inserting \eqref{E pot1}-\eqref{E pot3} into the electric boundary conditions  \footnote{The last equality comes from requiring the tangential derivatives of $V_i$ and $V_s$ be equal at the surface, which implies $V_i$ and $V_s$ are equal up to a constant which can be neglected.}:
\begin{align}
\epsilon \pd_n V_i - \pd_n V_s = (\epsilon-1)\hat{\b n}\cdot \b E_e|_S, \qquad V_i=V_s|_S. 
\end{align}
with $\pd_n=\hat{\b n}\cdot\nabla$.

For axisymmetric particles, $\hat{\b n}\cdot\hat{\bm\upphi}=0$ so 
\begin{align}
\hat{\b n}\cdot \b E_e = (\b r \times \hat{\b n})\cdot\nabla U = \frac{\hat{\b n}\cdot\hat{\bm\uptheta}}{\sin\theta}\frac{\pd U}{\pd\phi}
\end{align} 
which can be obtained directly from the component-wise expressions for the vector spherical harmonics (see Ref.~\cite{2002Mishchenko} App. C). We can obtain expressions for $\b T^{21}$ by solving for $V_s$ to obtain $\b q$ in terms of $\b a$, and comparing this solution with the matrix expression \eqref{TR}. This approach can also be used for $\b R^{21}$.

\subsection{Analytic expressions for spheroids}
\label{secAnalytic}

Spheroidal particles are a special case where there exists a full analytic solution in spheroidal coordinates. This provides a means to find analytic expressions for the entire $\b T$, $\b P$, and $\b Q$ matrices.
We follow a similar approach to Ref.~\cite{TmatESA2017}: solve the boundary problem in terms of spheroidal harmonics and re-express this in terms of spherical harmonics by applying basis transformations.  
We want to solve for the potentials $V_i,V_s$, knowing $U$. Since $U$, $V_i$, $V_s$ satisfy Laplace's equation, we can express them as series of spheroidal harmonics:
\begin{align}
U&=E_0\sum_{n,m}A_n^m P_n^m(\xi)P_n^m(\eta)e^{im\phi}, \\
V_i&=E_0\sum_{n,m}B_n^m P_n^m(\xi)P_n^m(\eta)e^{im\phi}, \\
V_s&=E_0\sum_{n,m}C_n^m  Q_n^m(\xi)P_n^m(\eta)e^{im\phi}. 
\end{align}
In light of evaluating the boundary conditions, for a spheroid we have $\hat{\b n}=\hat{\bm\upxi}$, and
\begin{align}
\hat{\bm\upxi}\cdot\hat{\bm\uptheta}&=\frac{\sin\theta~\eta}{\sqrt{(\xi^2-\eta^2)(\xi^2-1)}}, \label{xidottheta}\\
\frac{\pd}{\pd n}&=\frac{1}{f}\sqrt{\frac{\xi^2-1}{\xi^2-\eta^2}}\frac{\pd}{\pd \xi}.
\end{align}
The benefit of using spheroidal harmonics is that we can simply equate the coefficients of $P_n^m(\eta)$ in the expansions. For the factor of $\eta$ in \eqref{xidottheta}, we use the following identity:
\begin{align}
\eta P_n^m(\eta)=\frac{(n-m+1)P_{n+1}^m(\eta)+(n+m)P_{n-1}^m(\eta)}{2n+1}
\end{align}
and re-index the sums so that all terms contain $P_n^m(\eta)$. Then by equating the coefficients we obtain
\begin{align}
&B_n^m=\frac{Q_n^m(\xi_0)}{P_n^m(\xi_0)}C_n^m, \\
&C_n^m=\frac{ifm (\epsilon-1)P_n^{-m}(\xi_0)}{1+(s^2-1)L_n^m(\xi_0)} \times \nonumber\\
&\left(\frac{n\!-\!m}{2n\!-\!1}P_{n-1}^m(\xi_0)A_{n-1}^m + 
\frac{n\!+\!m\!+\!1}{2n\!+\!3}P_{n+1}^m(\xi_0)A_{n+1}^m \right).
\end{align}

Now we must express this solution on a spherical harmonic basis. 
The relevant relationships between the spherical and spheroidal harmonics are
\begin{align}
\left(\frac{r}{f}\right)^nP_n^m(\cos\theta)&=\sum_{k=0}^n \alpha_{nk}^m P_k^m(\xi)P_k^m(\eta), \\
Q_n^m(\xi)P_n^m(\eta)&=\sum_{k=n}^\infty \beta_{nk}^m \left(\frac{f}{r}\right)^{k+1}P_k^m(\cos\theta),
\end{align}
where the coefficients $\alpha_{nk}^m, \beta_{nk}^m$ are given in Appendix \ref{AppExpansions}. By substituting these expressions into the potential and electric field expressions, we find that the series coefficients must satisfy
\begin{align}
A_p^m&=\sum_{k=p}^\infty\alpha_{kp}^m\frac{(k_o f)^k}{(2k+1)!!}\gamma_k^m a_k^m, \\
q_n^m&=\frac{1}{\gamma_n^m}\frac{ik_o(k_of)^{p+1}}{n(2n-1)!!}\sum_{p=|m|}^n\bar\beta_{pn}^mC_p^m.
\end{align}
Combining these with the relationship between $C_n^m$ and $A_n^m$, we obtain an expression relating $\b q$ and $\b a$ which can be compared to \eqref{TR} to obtain $\b T^{21}$. We can also follow a similar derivation and obtain the quasistatic limit of $\b R^{21}$. We can then obtain expressions for $\b P^{21}$ and $\b Q^{21}$ from their matrix relationships to $\b T$ and $\b R$. We have $\b Q^{21}=-\b Q^{22}\b R^{21}\b Q^{11}$, which comes from the blockwise matrix inverse formula. Similarly, we can find  $\b R^{12}=-\b R^{11}\b Q^{12}\b R^{22}$ and $\b P^{12}=-\b T^{21}\b Q^{11}-\b T^{22}\b Q^{21}$. Below we summarise the results for all matrices:
\begin{widetext}
\begin{align}
T^{21(0)}_{nk|m}=&-(s^2-1)\frac{B_n^m B_k^m}{k+1}(k_of)^{n+k+2}(-)^m m\sum_{p=|m|}^{\min(n,k+1)}e_{np}~e_{k+1,p}~
\frac{ P_p^{-m}(\xi_0) }{1+(s^2-1)L_p^m(\xi_0)} \times \nonumber \\ 
&\frac{(p+m)(k+p+2)P_{p-1}^m(\xi_0)+(p-m+1)(k-p+1)P_{p+1}^m(\xi_0)}{(k-p+1)!!(k+p+2)!!(n-p)!!(n+p+1)!!},\\ \nonumber
\\
R^{21(0)}_{nk|m}=&-(s^2-1)\frac{B_k^m}{B_n^m}\frac{im(k_of)^{k-n+1}}{s^{n-1}(k+1)}
\sum_{p=n}^{k+1}e_{np}~e_{k+1,p}~
\frac{Q_p^{-m}(\xi_0)}{1+(s^2-1)L_p^m(\xi_0)} \times \nonumber \\ 
&(-)^{(n-p)/2+m}(p+n-1)!!\frac{(p+m)(k+p+2)P_{p-1}^m(\xi_0)+(p-m+1)(k-p+1)P_{p+1}^m(\xi_0)}{(k-p+1)!!(k+p+2)!!(p-n)!!},\\ \nonumber
\\
R^{12(0)}_{nk|m}=&-(s^2-1)\frac{B_k^m}{B_n^m}\frac{im(k_of)^{k-n+1}}{s^n n}
\sum_{p=n}^{k+1}e_{np}~e_{k+1,p}~
Q_p^m(\xi_0)\times \nonumber \\ 
&\frac{(-)^{(n-p)/2}(p+n-1)!!}{(k-p+1)!!(k+p+2)!!(p-n)!!}\left[\frac{(p-m)(k+p+2)P_{p-1}^{-m}(\xi_0)}{1+(s^2-1)L_{p-1}^m(\xi_0)}+\frac{(p+m+1)(k-p+1)P_{p+1}^{-m}(\xi_0)}{1+(s^2-1)L_{p+1}^m(\xi_0)}\right],\\ \nonumber
\\
Q^{21(0)}_{nk|m}=&-(s^2-1)s^k\frac{B_k^m}{B_n^m}\frac{im(k_of)^{k-n+1}}{k+1}
\sum_{p=n}^{k+1}e_{np}~e_{k+1,p}~
Q_p^{-m}(\xi_0)\times \nonumber \\ 
&(-)^{(n-p)/2+m}(p+n-1)!!\frac{(p+m)(k+p+2)P_{p-1}^m(\xi_0)+(p-m+1)(k-p+1)P_{p+1}^m(\xi_0)}{(k-p+1)!!(k+p+2)!!(p-n)!!},\\ \nonumber
\\
P^{21(0)}_{nk|m}=&(s^2-1)s^k\frac{B_n^m B_k^m}{k+1}(k_of)^{n+k+2}(-)^mm\sum_{p=|m|}^{\min(n,k+1)}e_{np}~e_{k+1,p}~
P_p^{-m}(\xi_0) \times \nonumber \\ 
&\frac{(p+m)(k+p+2)P_{p-1}^m(\xi_0)+(p-m+1)(k-p+1)P_{p+1}^m(\xi_0)}{(k-p+1)!!(k+p+2)!!(n-p)!!(n+p+1)!!}.
\end{align}
\end{widetext}
The expressions for $\b P^{21}$, $\b Q^{21}$ and $\b R^{12}$ were simplified using the following identity: 
\begin{align}
\sum_{q=p}^r e_{qr}~ \frac{(-)^{(r-q)/2}(r+q-1)!!}{(q-p)!!(q+p+1)!!(r-q)!!}=\frac{\delta_{pr}}{2r+1} \label{ID}
\end{align}
which  can be obtained by combining the expansions between spherical and spheroidal harmonics \eqref{PvsPP} and \eqref{PPvsP} and noting their orthogonality.

Note that the lower triangular parts of $\b Q^{12}$, $\b Q^{21}$, $\b R^{12}$, $\b R^{21}$ are zero within this
quasistatic approximation.
$\b P^{12}$ and $\b Q^{12}$ can moreover be obtained through:
\begin{align}
P_{nk|m}^{12(0)}&=\frac{1}{s}\frac{k+1}{n+1}P_{nk|m}^{21(0)}, \label{P21sym}\\
Q_{nk|m}^{12(0)}&=\frac{1}{s}\frac{k+1}{n}Q_{nk|m}^{21(0)} \qquad n\leq k+1 \label{Q21sym}. 
\end{align}
which can be derived for a general axisymmetric scatterer from the integral expressions given in Refs.~\cite{somerville2011simplified,TmatESA2017}.

\section{Quasistatic limit for $\b T^{11}$}
\label{secT11}
This block determines the scattered magnetic multipole field induced by an incident magnetic multipole field. For non-magnetic particles, this matrix is zero in the static case and only arises from non-zero frequency interactions.  
We can obtain the matrix elements using a similar method to that for $\b T^{21}$, this time formulating the problem in terms of magnetic fields.

\subsection{General formulation}

Following the approach for $\b T^{21}$, the matrix $\b T^{11}$ can be found by considering an incident field of magnetic multipoles, but here the spherical vector wave functions must be expanded to second order:
\begin{align}
\b H_e &= \sum_{n=m}^\infty a_{nm} [ \b {RgN}_{nm}^{(0)}(k_o\b r) + \b {RgN}_{nm}^{(2)}(k_o\b r) ], \\
\b H_i &= s \sum_{n=m}^\infty c_{nm}^{(0)} [ \b {RgN}_{nm}^{(0)}(k_i\b r) + \b {RgN}_{nm}^{(2)}(k_i\b r) ] \nonumber\\ 
&\qquad\quad+ c_{nm}^{(2)}\b {RgN}_{nm}^{(0)}(k_i\b r) + d_{nm} \b {RgM}_{nm}^{(0)}(k_i\b r), \label{Hi11}\\
\b H_s &= \sum_{n=m}^\infty p_{nm} \b N_{nm}^{(0)}(k_o\b r) + q_{nm} \b M_{nm}^{(0)}(k_o\b r).
\end{align}
This approach differs for the cases $m=0$ and $m\neq0$, since the off diagonal $T$-matrix blocks are zero for $m=0$. 
To lowest order, $c_{nm}^{(0)}=s^{-n}a_{nm}$ as in \eqref{R11},  and the leftmost terms in $\b H_e$ and $\b H_i$ are identical. $d_{nm}$, $p_{nm}$ and $q_{nm}$ are all kept to lowest order only. The problem is to solve for $p_{nm}$ and $c_{nm}^{(2)}$.

To express $\b {RgN}_{nm}^{(2)}$ with harmonic functions, we note that for a solution of the Helmholtz equation \cite{gumerov2007scalar}
\begin{align}
\nabla\times\nabla\times(\b r\psi) = \nabla [(1+r\pd_r)\psi] + \b r k^2\psi.
\end{align}
Then we express the incident field to second order as:
\begin{align}
\b H_e = \b H_e^{(0)} + \nabla (k_o^2r^2U_e^\nabla) +  \b r k_o^2 U_e^r \label{He2}
\end{align}
where both $U_e^\nabla$ and $U_e^r$ satisfy Laplace's equation. $\b H_e^{(0)}$ does not interact with the spheroid so may be left in vector form.
For the other fields we may write
\begin{align}
\b H_i &= \b H_e^{(0)} + s^2\big[\nabla (k_o^2r^2U_e^\nabla) +  \b r k_o^2 U_e^r\big] -\nabla U_i^N \nonumber\\&\hspace{5cm}
+ \nabla\times(\b r U_i^M), \\
\b H_s &= -\nabla U_s^N + \nabla\times(\b r U_s^M).
\end{align} 
As in \eqref{Hi11}, the first three terms of $\b H_i$ on the right hand side are known, while $U_i^M$ and $U_s^M$ are determined from $\b R^{21}$ and $\b T^{21}$. This leaves us to determine the two potentials $U_i^N$ and $U_s^N$. For this problem it appears most straightforward to apply the two magnetic field boundary conditions only, since for $m=0$ the condition for $\b{\hat{n}} \cdot \b E$ is redundant.

\subsection{Prolate spheroids, $m=0$}

For $m=0$, we assume an axially symmetric incident field. $\b T^{21}$ and $\b T^{12}$ are both zero and the problem is decoupled from any interactions of electric multipoles. This means $U_i^M=U_s^M=0$. Since $\hat{\bm\upphi}\cdot\b H=0$, the boundary conditions are $\hat{\bm\upxi}\cdot\b H_i=\hat{\bm\upxi}\cdot\b H_o$ and $\hat{\bm\upeta}\cdot \b H_i = \hat{\bm\upeta}\cdot \b H_o$, or in terms of the potentials:
\begin{align}
(s^2\!-\!1)k_o^2f^2\big[ \pd_\xi[(\xi^2\!+\!\eta^2\!-\!1)U_e^\nabla ] -\xi U_e^r \big] + \pd_\xi U_i^N = \pd_\xi U_s^N, \\
(s^2\!-\!1)k_o^2f^2\big[ \pd_\eta[(\xi^2\!+\!\eta^2\!-\!1)U_e^\nabla ] -\eta U_e^r \big] + \pd_\eta U_i^N = \pd_\eta U_s^N.
\end{align}
For the second boundary condition it is convenient to integrate over $\eta$. 
Then we expand the fields as series of spheroidal harmonics, apply recurrence identities for the Legendre polynomials, and re-index the summations to express the $\eta$ dependence of each term in the series as $P_n(\eta)$. 
There is a lot of algebra so we skip to the final result:
\begin{widetext}
\begin{align}
T_{nk|0}^{11(0)} =& -i(s^2-1)(\xi_0^2-1)(k_of)^{n+k+3}\frac{B_n^0B_k^0}{k+1}\sum_{p=0}^{\min(n,k)}
\frac{e_{nk}e_{np}}{(n-p)!!(n+p+1)!!(k-p)!!(k+p+1)!!}~ \nonumber\\
&\times\Bigg\{ (2p+1)\left(\frac{P_pP_p'}{(2p+3)(2p-1)}-\frac{k}{2k+3}\xi_0P_pP_p \right) -\frac{k-p}{k+p+3}\frac{p+2}{2p+3}P_{p+2}P_p' + \frac{n-p}{n+p+3}\frac{p+1}{2p+3}P_pP_{p+2}' \nonumber\\
&  - \frac{(k+3)(k-n)(p+1)(p+2)}{2(2k+3)(n+p+3)(k+p+3)}[P_pP_{p+2}'-P_{p+2}P_p']
\Bigg\}
\end{align}
\end{widetext}
where $P_p\equiv P_p(\xi_0)$.
Unlike the other T-matrix blocks, this has no log terms or singular points -- it is a polynomial in $\xi_0$.
Despite its appearance, one can check numerically that this expression is actually symmetric about $n$ and $k$, as it should be.
None of the terms individually are symmetric, making it hard to recognise a symmetric form of this expression. This suggests there could be simpler approach to obtaining the matrix, maybe where the 2$^\mathrm{nd}$ order fields are split differently to \eqref{He2} or a combination of one electric and one magnetic boundary condition could be applied instead.

\subsection{Prolate spheroids, $m\neq0$}

For $m\neq0$, the problem has the additional complication of coupling from the electric multipoles induced in both the internal and scattered fields. 
In this case the boundary condition on $\hat{\bm\upeta}\cdot\b H$  is too complicated, but the condition for $\hat{\bm\upphi}\cdot\b H$ is non-zero and manageable, so we have
\begin{align}
(s^2-1)k_o^2f^2\big[ \pd_\xi[(\xi^2+\eta^2-1)U_e^\nabla ] -\xi U_e^r \big]& \nonumber\\
+  \frac{\eta f}{\xi^2-1}\pd_\phi[U_s^M-U_i^M] =& \pd_\xi [U_s^N - U_i^N], \\
(s^2-1)k_o^2r^2 U_e^\nabla  - r\sin\theta\pd_\theta [ U_s^M - U_i^M ] = & \pd_\phi [ U_s^N - U_i^N ] .
\end{align}
The derivative $\pd_\theta$ can be applied directly to the spherical harmonics, which splits them into two different orders, adding another layer  of complication.
All potentials are harmonic and should be expanded on a basis of spheroidal harmonics, and then related to their corresponding expansion in spherical wave functions. The series coefficients $d_n^m$, $q_n^m$ for $U_i^M$, $U_s^M$ are given by $\b R^{21}$ and $\b T^{21}$. The final result is:
\begin{widetext}
\begin{align}
T_{nk|m}^{11(0)} =& \frac{-i(\xi_0^2-1)(k_of)^{n+k+3}(n-m)!}{\gamma_n^m n(2n-1)!!}\sum_{p=m}^n\frac{e_{nk}e_{np}}{(n-p)!!(n+p+1)!!} \Bigg\{ \frac{(s^2-1)\gamma_k^m}{(2k+3)!!}~\alpha_{kp}^m \nonumber\\
&\times \left[ \frac{(k-n)(p-m+1)(p-m+2)(k+3)}{2(2p+1)(n+p+3)(k+p+3)}[P_p^mP_{p+2}^{m\prime}-P_{p+2}^mP_p^{m\prime}] - k\xi_0P_p^mP_p^m \right] \nonumber \\
+& \sum_{q=\max(p-1,m)}^{k+1}\frac{i\gamma_q^m(k_of)^{q-k-1}s^{q+1}}{(2q+1)!!} R_{qk|m}^{21}\nonumber \\
 & \times \left[ -\left(\frac{mP_p^m}{\xi_0^2-1}+\xi_0P_p^{m\prime}\frac{q+1}{m}\right)\left(\frac{p-m}{2p-1}\alpha_{q,p-1}^mP_{p-1}^m + \frac{p+m+1}{2p+3}\alpha_{q,p+1}P_{p+1}^m\right) + \frac{q-m+1}{m}\alpha_{q+1,p}^mP_p^m P_p^{m\prime} \right] 
\nonumber\\
+&  \sum_{q=m}^{p+1}\frac{\gamma_q^m(2q-1)!!}{(k_of)^{q+k+2}} T_{qk|m}^{21} \nonumber \\
 & \times \left[ \left(\frac{mP_p^m}{\xi_0^2-1}-\xi_0P_p^{m\prime}\frac{q}{m}\right)\left(\frac{p-m}{2p-1}\beta_{q,p-1}^mQ_{p-1}^m + \frac{p+m+1}{2p+3}\beta_{q,p+1}Q_{p+1}^m\right)  + \frac{q+m}{m}\beta_{q-1,p}^mQ_p^m P_p^{m\prime} \right] 
\Bigg\}.
\end{align}
\end{widetext}
Where $\alpha_{nk}^m$ and $\beta_{nk}^m$ are the coefficients in the expansions \eqref{PvsPP} and \eqref{PvsQP} in the appendix. 
Again the matrix is symmetric despite its appearance, and it is likely that simplified expressions could be found.

\section{Discussion}
\label{disc}

\subsection{Evaluating and checking expressions} \label{evaluating}
For convenience, Matlab codes to evaluate the quasistatic matrices are attached as supplementary material. 

For oblate spheroids, as mentioned earlier, all expressions for the matrices in this paper can be used as they are. The Legendre functions should be defined from their definition ```off the cut'', ie. all complex space except on the real line from $-1$ to $1$.
Matlab codes to evaluate these Legendre functions are also provided. 

\begin{figure*}
\includegraphics[scale=.75]{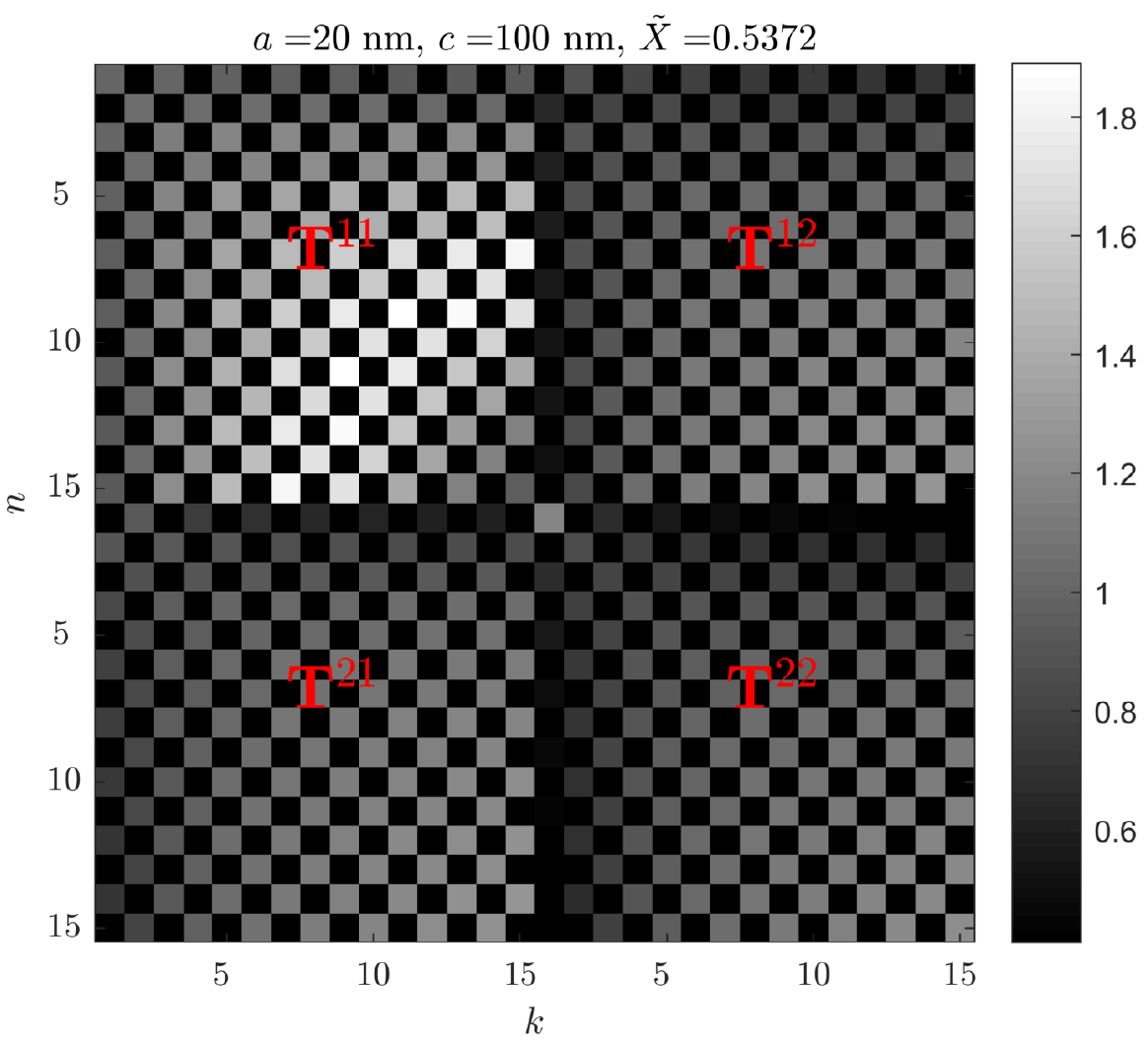} 
\includegraphics[scale=.75]{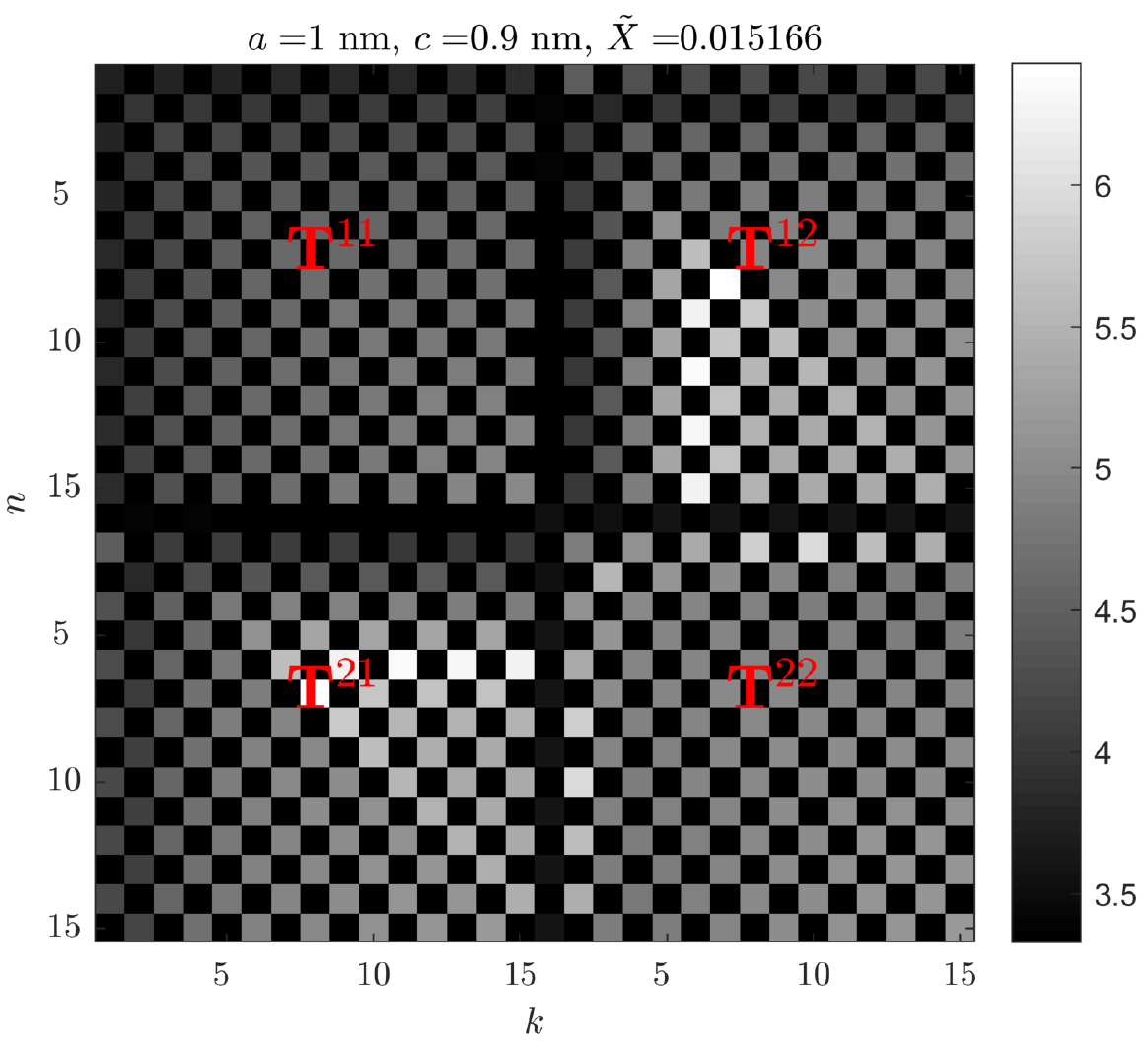} 
\caption{$-\log_{10}$(relative error) of the quasistatic $T$-matrix expressions compared to the exact solutions, for a silver spheroid in water, with a wavelength of 400 nm, $(\epsilon=-6.4572+0.2993i)$} \label{err plots}
\end{figure*}

All the obtained expressions were checked against the exact $T$-matrix results, which can be computed to a high accuracy
\cite{JQSRT2015,2016SMARTIES}, and the relative error is plotted in figure \ref{err plots}. The radiative correction was also applied to the quasi-static $T$-matrix: $\b T\rightarrow \b T(\b I-\b T)^{-1}$ \cite{2013LeRuPRA}, which had a noticeable increase in accuracy for larger particles (for example the spheroid on the right plot). Results are similar to the error plots for $\b T^{22}$ presented in \cite{TmatESA2017}.
This numerical check provides independent confirmation of the validity of these analytic formulae. The size parameter is $\tilde{X}=k_1r_\text{eq}=k_1\sqrt[3]{c^2a}$ where $r_\text{eq}$ is the radius of the volume equivalent sphere.

The accuracy generally improves as the size parameter $\tilde{X}$ decreases, and the approximations appear to somewhat favour low aspect ratios.
Accuracy increases modestly with order, and tends to be more accurate for $\epsilon$ with positive real part. 

\subsection{Depolarization factors}
\label{secLnm}

The $T$- and $R$-matrix expressions contain what we may call generalised depolarisation factors:
\begin{align}
L_n^m(\xi)=(\xi^2-1)P_n^{m\prime}(\xi)Q_n^{-m}(\xi), \label{Lnm}
\end{align}
which reduce to the well known dipolar depolarisation factors $L_x,L_y,L_z$ for $n=1$, and obey the sum rule
(see appendix for proof):
\begin{align}
\sum_{m=-n}^n L_n^m=n. \label{sum rule}
\end{align}
For $n=1$ this is equivalent to $L_x+L_y+L_z=1$. 

We can also find integral expressions for the depolarisation factors by comparison with the EBCM.
In the quasistatic limit of the EBCM for axisymmetric particles the diagonal elements of $\b Q^{22}$ may be expressed as (after some manipulation)
\begin{align}
Q_{nn|m}^{22}
&=s^{n-1}\bigg\{1+(s^2-1)\bigg[\frac{n}{2n+1}-\frac{(-)^m}{2}\times\nonumber\\&~~\int_0^\pi \d \theta \sin\theta P_n^{-m}(\cos\theta)\frac{\d P_n^m(\cos\theta)}{\d\theta}\frac{1}{r(\theta)}\frac{\d r(\theta)}{\d\theta}\bigg]\bigg\}.\label{Qint}
\end{align}
where $r(\theta)$ defines the surface of the scatterer. Eq.~\eqref{Qint} reduces to the approximate (if the internal field is nearly uniform) dipolar responses in \cite{il2011rayleigh}.
As $r(\theta)$ becomes constant (i.e. for a sphere), the integral disappears.

For spheroids
\begin{align}
r(\theta)=c\sqrt{\frac{\xi_0^2-1}{\xi_0^2-\cos^2\theta}} \Rightarrow \frac{1}{r}\frac{\d r}{\d\theta}=-\frac{\sin\theta\cos\theta}{\xi_0^2-\cos^2\theta}. 
\end{align}
and from \eqref{esQ} we have for the diagonal
\begin{align}
Q_{nn|m}^{22}=s^{n-1}[1+(s^2-1)L_n^m]. \label{QL}
\end{align}
Comparing \eqref{QL} and \eqref{Qint} we find
\begin{align}
L_n^m=\frac{n}{2n+1}+\frac{(-)^m}{2}\int_0^\pi \d \theta  P_n^{-m}\frac{\d P_n^m}{\d\theta}\frac{\sin^2\theta\cos\theta}{\xi_0^2-\cos^2\theta}. \label{Lint}
\end{align}
This agrees numerically with \eqref{Lnm}, although analytically is not obvious. We have introduced \eqref{Lint} not as a practical device but as an extension of the integral expression in \cite{il2011rayleigh} and to highlight the connection with the EBCM.

\subsection{Quasistatic plasmon resonances of spheroids}
\label{secRes}

In all matrix elements expressions, we see that the expression $1+(\epsilon-1)L_n^m(\xi_0)$ appears as a denominator.
For metallic scatterers, the real part of $\epsilon$ is negative and this denominator can approach zero for a certain wavelength if losses (imaginary part) are small.
This result in a very strong optical response commonly referred to as a localized surface plasmon resonance (LSPR) of the nanoparticle \cite{2009book}.
For a sphere, these resonances occur for $\Re (\epsilon) =-2$ for the dipolar resonance and $\Re (\epsilon) =-(n+1)/n$ for $n$-multipolar resonances.

The generalized depolarization factors allow us to define and study these resonances in spheroids.
They occur for:
\begin{align}
\Re \left(\epsilon_{res}\right)
=1-\frac{1}{L_n^m(\xi_0)}\label{res}
\end{align}
for all $n\geq 1$, $m\geq 0$. For $n=1$, this reduces to the known dipolar LSPR of a spheroid \cite{2009book}, where excitation along $z$ corresponding to $L_1^0$, and along $x$ or $y$ corresponding to $L_1^1$.

The elements $T_{nk|m}^{ij}$ with $n\leq2$ or $k\leq2$ have just one resonance, and many elements share the same resonance condition, for example $T_{11|1}^{11}, T_{21|1}^{21}, T_{22|1}^{22}$ all resonate at $1+(\epsilon-1)L_2^1(\xi_0)=0$.
These resonances are well known from the solution of the scattering problem in spheroidal coordinates \cite{knipp1992classical}, but here the resonances are associated with their excitations from \textit{spherical} multipoles through the $T$-matrix. 
For small aspect ratios, the resonances are split relative to the spherical case, and for higher orders $n$, there are more splittings (one for each $m\leq n$), but the shifts are small.

In the spherical limit, $\xi_0\rightarrow\infty$, and the limits of the Legendre functions are (\cite{tables2014}, 8.776):
\begin{align}
\lim_{\xi\rightarrow\infty}P_n^m(\xi)&=\frac{(2n-1)!!}{(n-m)!}\xi^n \\
\lim_{\xi\rightarrow\infty}Q_n^m(\xi)&=(-)^m\frac{(n+m)!}{(2n+1)!!}\xi^{-n-1}\\
\lim_{\xi\rightarrow\infty} L_n^m&=\frac{n}{2n+1}
\end{align}
which lead to the small sphere resonance conditions, $\Re(\epsilon_{res})=-(n+1)/n$, as expected.

\section{Conclusion}

We have provided an approach to find the quasistatic  limit of $\b T$ for any axisymmetric particle, and in the case of spheroids, this approach leads to analytic expressions.

For non-magnetic particles, the magnetic multipole field does not interact with the object in the static limit. This means that this interaction cannot be seen without considering at least the lowest two orders of the spherical wave functions. This is exactly the case for a sphere, where the $T$-matrix reduces to the electric and magnetic Mie susceptibilities. The quasistatic limit of the electric susceptibilities is obtained from an electrostatics problem, while in this limit the magnetic susceptibilities are zero. For magnetic particles however, $\b T^{11}$ is non-zero to the lowest order and could be obtained from a magnetostatics problem, with very similar formalism to the electrostatics problem for $\b T^{22}$.

 
In a recent paper \cite{SpheroidApproximation2018} the $T$-matrix was found to 3$^\mathrm{rd}$ lowest order, i.e. $\cO(X^6)$, which involves only up to multipolarity $n=3$. The results were derived by direct Taylor expansion of the EBCM, and are particularly relevant in the context of plane wave scattering. In contrast, here we have found the lowest non-zero order of the individual elements, by considering a point source excitation where every $T$-matrix element is equally important. Some of the results of these two approaches coincide, in particular the lowest orders of $T_{11|m}^{22}$, $T_{11|m}^{11}$, $T_{22|m}^{22}$, $T_{13|m}^{22}$, $T_{12|1}^{21}$, $T_{21|1}^{21}$ (and their symmetric counterparts) for $m=0,1,2$. These results were used to confirm and simplify some of the ECBM-derived expressions.

We believe these analytic expressions will be useful in fundamental studies of the $T$-matrix method, for example in relation with the Rayleigh hypothesis and analysis of quasistatic resonances.

\section*{Acknowledgments}
We acknowledge the support of the Royal Society of New Zealand (RSNZ) through a Marsden Grant (ECLR).

\appendix

\section{Definitions of Legendre functions}
There exist different definitions of the Legendre functions $P_n^m(x), Q_n^m(x)$ for $x$ real, $|x|<1$ (the branch cut) and off the cut. First of all, the Legendre polynomials are defined as (which applies on and off the cut)
\begin{align}
P_n(x)=\frac{1}{2^nn!}\frac{\d^n}{\d x^n}(x-1)^n.
\end{align}

For the spherical harmonics (where $\cos\theta$ is on the cut):
\begin{align}
P_n^m(\cos\theta)=\sin^m\theta\bigg(\frac{\d}{\d \cos\theta}\bigg)^mP_n(\cos\theta). 
\end{align}
Some authors multiply by $(-)^m$ in their definition.
For the spheroidal harmonics, the coordinate $\xi$ is off the branch cut for both prolate and oblate spheroidal coordinates. Here the Legendre functions of the second kind are
\begin{align}
Q_n(\xi)=\frac{1}{2}\int_{-1}^1\frac{P_n(t)}{\xi-t}\d t.
\end{align}
For $m>0$ this manuscript uses the definitions
\begin{align}
P_n^m(\xi)&=(\xi+1)^{m/2}(\xi-1)^{m/2}\frac{\d^m}{\d \xi^m}P_n(\xi), \\
Q_n^m(\xi)&=(\xi+1)^{m/2}(\xi-1)^{m/2}\frac{\d^m}{\d \xi^m}Q_n(\xi).
\end{align}
The factors $(\xi+1)^{m/2}(\xi-1)^{m/2}$ have not been combined into $(\xi^2-1)^{m/2}$ in order to give the correct results for all complex $\xi$ off the cut. These definitions coincide with the general definitions in terms of hypergeometric functions, which are implemented in Maple as \texttt{LegendreQ(n,m,x)} and in Mathematica as \texttt{LegendreQ[n,m,3,x]}. 


Moreover, for negative order, we have:
\begin{align}
P_n^{-m}=(-)^m\frac{(n-m)!}{(n+m)!}P_n^m
\end{align}
and similarly for $Q_n^m$.

Finally, their derivatives can be evaluated as
\begin{align}
\frac{\d P_n^m(\xi)}{\d \xi} =\frac{(n-m+1) P_{n+1}^m(\xi)-(n+1)\xi P_n^m(\xi)}{\xi^2-1}, \\
\frac{\d Q_n^m(\xi)}{\d \xi} =\frac{(n-m+1) Q_{n+1}^m(\xi)-(n+1)\xi Q_n^m(\xi)}{\xi^2-1}.
\end{align}

\section{Proof of sum rule for $L_n^m$}
First we re-express the sum rule \eqref{sum rule} as
\begin{align}
\sum_{m=-n}^n P_n^{-m\prime}(\xi)Q_n^m(\xi)=\frac{n}{\xi^2-1}. \label{sumrule-m}
\end{align}
We will use the result 
\begin{align}
\sum_m P_n^{-m}(\xi)Q_n^m(\xi)=Q_0(\xi),\label{additiontheoremQ}
\end{align}
which can be proved by integrating a special case of the ``addition theorem" \cite{arfken1999mathematical}:
\begin{align}
\sum_m (-)^mP_n^{-m}(x)P_n^m(x)=1 \label{additiontheorem}
\end{align} 
and using \cite{macrobert1952neumann} (note their $P_n^m(z>1)$ is out by $(-)^m$ from more recent definitions, including ours)
\begin{align}
\frac{1}{2}\int_{-1}^1 \frac{P_n^{-m}(x)P_n^m(x)}{\xi-x}\d x = (-)^mP_n^{-m}(\xi)Q_n^m(\xi)
\end{align}
for both the left and right hand sides of \eqref{additiontheorem}. \\
Then we may differentiate \eqref{additiontheoremQ} to obtain:
\begin{align}
\sum_m P_n^{-m\prime}(\xi)Q_n^m(\xi)+P_n^{-m}(\xi)Q_n^{m\prime}(\xi)=\frac{-1}{\xi^2-1}. \label{diffadditiontheoremQ}
\end{align}
We can then sum the Wronskian relation of the Legendre functions over $m$
\begin{align}
P_n^{-m\prime}(\xi)Q_n^m(\xi)-P_n^{-m}(\xi)Q_n^{m\prime}(\xi)&=\frac{-1}{\xi^2-1} \\
\Rightarrow \sum_m P_n^{-m\prime}(\xi)Q_n^m(\xi)-P_n^{-m}(\xi)Q_n^{m\prime}(\xi)&=-\frac{2n+1}{\xi^2-1}. \label{sumWronsk} 
\end{align}
Finally, \eqref{diffadditiontheoremQ} and \eqref{sumWronsk} may be combined and rearranged to obtain \eqref{sumrule-m}.

\section{Relationships between spherical and spheroidal solid harmonics}
\label{AppExpansions}

Below are the relations between spherical and spheroidal harmonics used throughout the manuscript. The azimuthal dependence $e^{\pm im\phi}$ is omitted since it is the same on both sides. Derivations can be found in \cite{Jansen2000,Antonov2002}.
\begin{align}
&P_n^m(\xi)P_n^m(\eta)=\frac{(n+m)!}{(n-m)!} \sum_{k=m}^n e_{nk} ~(-)^{(n-k)/2}\nonumber\\
&\quad\times\frac{(n+k-1)!!}{(n-k)!!(k+m)!}\left(\frac{r}{f}\right)^kP_k^m(\cos\theta) \label{PPvsP}
\\[0.6cm]
&\left(\frac{r}{f}\right)^nP_n^m(\cos\theta)=(n+m)!\sum_{k=m}^n e_{nk} \nonumber\\
&\quad\times\frac{(2k+1)}{(n-k)!!(n+k+1)!!}\frac{(k-m)!}{(k+m)!} P_k^m(\xi)P_k^m(\eta) \label{PvsPP}
\\[0.6cm]
&Q_n^m(\xi)P_n^m(\eta)=(-)^m\frac{(n+m)!}{(n-m)!}\sum_{k=n}^\infty e_{nk}~\nonumber\\
&\quad\times\frac{(k-m)!}{(k-n)!!(k+n+1)!!}\left(\frac{f}{r}\right)^{k+1}\!P_k^m(\cos\theta) \label{QPvsP}
\\[0.6cm]
&\left(\frac{f}{r}\right)^{n+1}\!P_n^m(\cos\theta)=\frac{1}{(n-m)!}\sum_{k=n}^\infty e_{nk} ~(-)^{(n-k)/2+m}\nonumber\\
&\quad\times\frac{(2k+1)(n+k-1)!!}{(k-n)!!}\frac{(k-m)!}{(k+m)!} Q_k^m(\xi)P_k^m(\eta) \label{PvsQP} 
\end{align}

These expressions can be written more concisely by defining the coefficients:
\begin{align}
\alpha_{nk}^m&=\frac{(n+m)!(2k+1)}{(n-k)!!(n+k+1)!!}\frac{(k-m)!}{(k+m)!}  \label{alpha} \\
\alpha_{nk}^m&=0 \qquad n<k ~\text{ or }~ n+k \text{ odd} \nonumber
\end{align}
\begin{align}
\beta_{nk}^m&=(-)^{(n-k)/2+m}\frac{(2k+1)(n+k-1)!!}{(n-m)!(k-n)!!}\frac{(k-m)!}{(k+m)!} . \label{beta}\\
\beta_{nk}^m&=0 \quad n>k \qquad n<k ~\text{ or }~ n+k \text{ odd}\nonumber
\end{align}

\bibliographystyle{elsarticle-num}
\bibliography{libraryH}
\end{document}